\documentclass{sig-alternate}
\usepackage{mathptmx} 

\usepackage{comment}
\usepackage{fancyhdr}
\usepackage[normalem]{ulem}
\usepackage[hyphens]{url}
\usepackage[sort,compress]{cite}
\usepackage[final]{microtype}
\usepackage{flushend}
\usepackage[bookmarks=true,breaklinks=true,letterpaper=true,colorlinks,linkcolor=black,citecolor=blue,urlcolor=black]{hyperref}

\newcommand{\microsubmissionnumber}{31}

\fancypagestyle{firstpage}{
  \fancyhf{}
  
  \fancyhead[C]{\vspace{15pt}\normalsize{MICRO 2017 Submission
      \textbf{\#\microsubmissionnumber} -- Confidential Draft -- Do NOT Distribute!!}} 
  \fancyfoot[C]{\thepage}
}

\title{Using Branch Predictors to Monitor Brain Activity\vspace{-12mm}} 
\author{Abhishek Bhattacharjee\\Department of Computer Science, Rutgers University}

\begin{document}
\maketitle
\pagestyle{plain}

\begin{abstract}

A key problem with neuroprostheses and brain monitoring interfaces is
that they need extreme energy efficiency. One way of lowering energy
is to use the low power modes available on the processors embedded in
these devices.  We present a technique to predict when neuronal
activity of interest is likely to occur, so that the processor can run
at nominal operating frequency at those times, and be placed in low
power modes otherwise. To achieve this, we discover that branch
predictors can also predict brain activity. By performing brain
surgeries on awake and anesthetized mice, we evaluate several branch
predictors and find that perceptron branch predictors can predict
cerebellar activity with accuracies as high as 85\%. Consequently, we
co-opt branch predictors to dictate when to transition between low
power and normal operating modes, saving as much as 59\% of processor
energy.

\end{abstract}

\section{Introduction}\label{sec:introduction}

Recent advances in invasive/non-invasive brain monitoring technologies
and neuroprostheses have begun she\-dding light on brain
function. Devices such as cochlear and retinal implants, as well as
emerging brain-machine interfaces for persons afflicted by spinal cord
injuries, motor neuron diseases, and locked-in syndrome are undergoing
rapid innovation \cite{graimann:brain, su:wireless, zanos:neurochip2,
  liu:machine, muhl:brain, ferreira:survey, mone:sensors}. This is
partly because the technologies used to probe and record neuronal
activity in vivo are fast improving -- we can currently monitor the
activity of hundreds of neurons simultaneously, and this number is
doubling approximately every seven years \cite{stevenson:how}. This
means that scientists can now study large-scale neuronal dynamics and
draw connections between their biology and higher-level cognition.

A natural consequence is that designers are beginning to integrate
embedded processors on neuroprostheses to achieve more sophisticated
computation than what was previously possible with the simple
micro-controllers and analog hardware traditionally used on these
devices \cite{angotzi:programmable, nguyen:closed, hirata:fully,
  liu:machine, mestais:wimagine, zanos:neurochip2, su:wireless,
  graimann:brain}. For example, embedded processors are beginning to
be used to perform sub-millisecond spike detection and sorting for
closed-loop experiments in which a stimulus is immediately delivered
to the brain whenever a specific neuron fires \cite{zanos:neurochip2,
  open-ephys}. Similarly, brain machine interfaces are replacing bulky
and inconvenient wired connections to large desktops with embedded
processors \cite{graimann:brain, palumbo:embedded, mercado:embedded,
  babu:embedded}.

These processors face a key obstacle -- they need to be energy
efficient. Consider the cerebellum, an important portion of the
hindbrain of all vertebrates. Recent studies use invasive brain
monitoring to record intracellular cerebellar neuronal activity
\cite{ozden:reliable, gauck:control, sullivan:invivo}. Invasive
monitoring implants must typically not exceed stringent 50-300mW power
budgets \cite{su:wireless, zanos:neurochip2, angotzi:programmable,
  nguyen:closed, hirata:fully, mestais:wimagine}. This is because
neural implants have small form factors and must therefore use the
limited lifetimes of their small batteries judiciously
\cite{zanos:neurochip2, su:wireless, angotzi:programmable,
  nguyen:closed, hirata:fully, mestais:wimagine}. Equally importantly,
stretching out battery lifetimes can reduce how often invasive
surgeries for battery replacement and/or recharging are
needed. Finally, power consumption must be kept low, as temperature
increases in excess of 1-2 degrees celcius can damage brain tissue
\cite{yarmolenko:thresholds, matsumi:thermal,
  wolf:thermal}. Unfortunately, the embedded processors used on
implants (typically from the ARM Cortex M line) can currently present
a barrier to energy efficiency in some systems, expending 35-50\% of
system energy \cite{su:wireless, zanos:neurochip2, kohler:flexible}.

A potential solution is to use the low power modes already available
on these processors \cite{kaxiras:computer, lefurgy:energy,
  herbert:dvfs, deng:multiscale, deng:coscale,
  deng:memscale}. Traditional energy management on server and mobile
systems balances the energy savings of low power modes with
performance degradation, by anticipating periods of time when
applications do not need certain resources or can afford slowdown
\cite{deng:memscale, deng:coscale, deng:multiscale, kaxiras:computer,
  bhattacharjee:thread, li:thrifty, meisner:dreamweaver,
  meisner:power, meisner:powernap}. Similar approaches are potentially
applicable to brain implants. Since embedded processors on implants
perform signal processing on neuronal spiking data, they could
theoretically be placed in low power mode in the absence of neuronal
firing and be brought back to nominal operation before neuronal
activity of interest. This presents the following problem -- how can
we predict when future neuronal spiking is likely to occur, both
accurately and efficiently?

In response, we observe the following. Architects have historically
implemented several perfor\-mance-critical micro\-architectural
structures to predict future program behavior. One such structure, the
bran\-ch predictor, is a natural candidate for neuronal prediction
too. Branch predictors assess whether a branch is likely to be taken
or not, and as it turns out, map well to the question of whether a
neuron fires or not at an instant in time. We study several branch
predictors and discover that the perceptron branch predictor
\cite{jimenez:perceptron, jimenez:fast, jimenez:piecewise, amant:high,
  jimenez:neural} can accurately predict future cerebellar neuronal
activity. We co-opt the perceptron predictor to not only predict
program behavior, but to also guide energy management of a cerebellar
monitoring implant. Our contributions are:

\vspace{2mm} {\noindent \bf \textcircled{1}} We evaluate well-known
dynamic hardware branch predictors, including Smith predictors
\cite{lee:branch, smith:study}, gshare \cite{mcfarling:gshare},
two-level adaptive predictors \cite{yeh:two}, and the perceptron
predictor \cite{jimenez:perceptron}. We perform surgeries on awake and
anesthetized mice to extract 26 minutes of neuronal spiking activity
from their cerebella and find that perceptron branch predictors are
particularly effective at predicting neuronal activity, with
accuracies as high as 85\%. The success of the perceptron predictor
can be attributed to the fact that it captures correlations amongst
longer histories of branches better than other approaches. This fits
well with cerebellar neuronal activity, where groups of neurons also
tend to have correlated activity \cite{ozden:reliable,
  sullivan:invivo}. These insights set the foundation for future
studies on using other branch predictors beyond the ones we study
\cite{seznec:analysis, jimenez:piecewise, seznec:tage,
  jimenez:optimized} for neuronal prediction.

\vspace{2mm} {\noindent \bf \textcircled{2}} We model a cerebellar
monitoring implant. Using architectural, RTL, and circuit modeling, we
use the branch predictor to not only predict branches but to also
guide energy management. We place the processor in idle low power mode
but leave the predictor on to predict brain activity. Wh\-en the
predictor anticipates interesting future cerebellar behavior, it
brings the processor back to normal operating mode (where the
predictor goes back to being a standard branch predictor).  Overall,
we save up to 59\% of processor energy.

\vspace{2mm} An important theme of this work is to ask -- since
machine learning techniques inspired by the brain have been distilled
into hardware predictors (e.g., like the perceptron branch predictor),
can we now close the loop and use such predictors to anticipate brain
activity and manage resources on neuroprostheses?  Our work is a first
step in answering this question. Ultimately, this approach can guide
not only management of energy, but also other scarce resources.

\section{Background}\label{sec:background}

\subsection{The Cerebellum}\label{subsec:the-cerebellum}

\begin{figure}[t]
\centering
{
\begin{minipage}[t]{0.48\textwidth}
\centering
\vspace{-4mm}
\epsfig{file=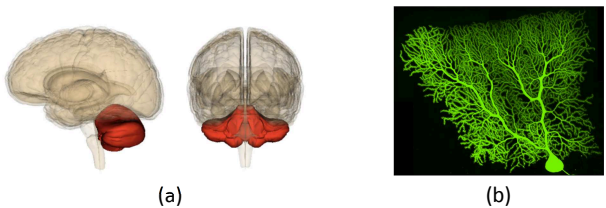, scale=0.36, angle=90}
\vspace{-50mm}
\caption{\small (a) The cerebellum, shown in red, is located behind
  the top of the brain stem and has two hemispheres
  \cite{cerebellum:image}; (b) a major cerebellar neuron is the
  Purkinje neuron, imaged here from a mouse brain
  \cite{purkinje:image}.}
\vspace{-4mm}
\label{fig:cerebellum-purkinje-general}
\end{minipage}
}
\end{figure}

The cerebellum (Latin for ``little brain'') affects motor control,
language, attention, and regulates fear and pleasure responses
\cite{ozden:reliable, gauck:control, sullivan:invivo, tank:spatially}.
It receives input from the sensory systems of the spinal cord and from
other parts of the brain, integrating them to fine-tune motor
activity. Cerebellar damage leads to movement, equilibrium, and motor
learning disorders. Cerebellar damage may also play a role in
hallucination and psychosis \cite{bielawski:psychosis, picard:role,
  kaloshi:visual}. Figure \ref{fig:cerebellum-purkinje-general} shows
the location of the cerebellum in the human brain and an in vivo image
of one of its major neuron types, the Purkinje neuron. Our goal is to
enable energy-efficient recording of Purkinje activity.

\begin{figure}[t]
\centering
{
\begin{minipage}[t]{0.48\textwidth}
\centering
\vspace{-4mm}
\epsfig{file=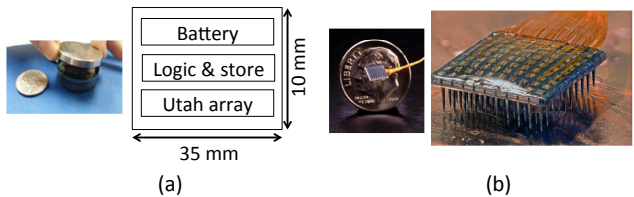, scale=0.36, angle=90}
\vspace{-50mm}
\caption{\small (a) Block diagram of cerebellar implant (dimensions
  not drawn to scale) and compared against a coin \cite{su:wireless};
  (b) the Utah array is used to collect intracellular Purkinje
  recordings \cite{utah:image1, utah:image2}.}
\vspace{-6mm}
\label{fig:implant-general}
\end{minipage}
}
\end{figure}

\subsection{Cerebellar Monitoring Implants}\label{subsec:cerebellar-monitoring-omplants}

Figure \ref{fig:implant-general} shows a cerebellar implant, and is
typical of recent neural implants \cite{su:wireless,
  zanos:neurochip2}. Most neural implants are small and placed in
containers embedded via a hole excised in the skull, from where they
probe brain tissue. Figure \ref{fig:implant-general} shows that
implants typically have the following components:

\vspace{2mm}{\noindent \bf Microelectrode array:} In vivo neuronal
activity is picked up using microelectrode arrays, which have improved
rapidly in recent years \cite{stevenson:how}. Many implants, including
our target system, use Utah arrays made up of several tens of
conductive silicon needles that capture intracellular recordings
\cite{kelly:comparison, suner:reliability, nordhausen:single}. Utah
arrays are widely used because of their high signal fidelity,
robustness, and relative ease of use.

\vspace{2mm}{\noindent \bf Logic and storage:} Neuronal activity
recorded by the Utah array is boosted by analog amplifier arrays
connected to anal\-og-to-digital converters (ADCs). While ADC designs
vary, 16-channel ADCs produce good signal integrity without excessive
energy usage \cite{su:wireless, angotzi:programmable}. ADCs route
amplified data to memory locations in LPDDR DRAM. Further, flash
memory cards are used to store neuronal data
\cite{mestais:wimagine}. Since GBs of neuronal activity data can be
generated in just tens of minutes of recording, most implants use a
wireless communication link (typically a GHz RF link) to transmit data
to a desktop system with sufficient storage for all the data being
recorded. Finally, embedded processors (e.g., energy-efficient ARM
Cortex M cores) are integrated on these implants \cite{su:wireless,
  zanos:neurochip2, angotzi:programmable, mestais:wimagine}. Our
studies focus on an implant with an embedded processor with similar
microarchitecture to the Cortex M7 (see Sec. \ref{sec:methodology} for
details). These processors run at 200-300 MHz, but maintain two
low-power modes to turn either the processor clock off (to roughly
halve processor power consumption) or turn off DRAM and flash too (to
lower system power consumption by an order of magnitude)
\cite{cooreman:power}.

\vspace{2mm}{\noindent \bf Battery:} Designers commonly use 3.7V
batteries to power the implant. Ideally, we want to run implants
designed for mice for days or weeks. For primates, we want to push
battery lifetimes to months and years. Naturally, the longer the
lifetime, the better, since surgeries to replace batteries can be
avoided \cite{graimann:brain}. Wireless charging can reduce the need
for surgeries; nevertheless, energy efficiency remains important
because neural implants must not raise temperature beyond 1-2 degrees
celcius to prevent tissue damage \cite{yarmolenko:thresholds,
  wolf:thermal, matsumi:thermal}. As a result, designers aim to run
implants with power budgets of 50-100mW, occasionally permitting
higher budgets up to 300mW, if they are for brief periods of time
\cite{su:wireless, zanos:neurochip2, angotzi:programmable,
  mestais:wimagine}.

\section{Motivation}\label{sec:motivation}

Utah arrays and analog components are designed judiciously and turned
off when their services are deemed unnecessary\cite{su:wireless,
  zanos:neurochip2, angotzi:programmable,
  mestais:wimagine}. Consequently, today's embedded processors
sometimes use 35-50\% of system-wide energy, with components like
DRAM, flash, and analog circuits consuming the rest
\cite{su:wireless}. We attack this bottleneck using low power
modes. To do this, we answer several important questions:

\vspace{2mm}{\noindent \bf What low power modes do we use?}  We study
signal processing workloads that execute whenever there is neuronal
activity of interest. Correct operation requires memory and
architectural state to be preserved during low power mode.  Deep low
power modes that lose state are hence infeasible. On ARM Cortex
processors, with the two low power modes detailed in
Sec.~\ref{sec:background}, this means that we are restricted to using
modes that turn off the processor and caches, but not DRAM or flash
memory. In the future, as Cortex M7 processors adopt stateful low
power modes for DRAM and flash memory \cite{deng:coscale,
  deng:memscale}, we anticipate using them too.

\begin{figure}[t]
\centering
{
\begin{minipage}[t]{0.48\textwidth}
\centering
\vspace{-4mm}
\epsfig{file=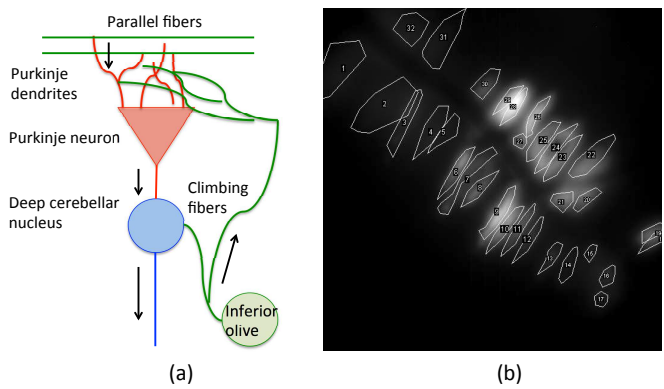, scale=0.36, angle=90}
\vspace{-26mm}
\caption{\small (a) Purkinje neurons are activated by the inferior
  olive and parallel fibers; and (b) synchronized activity (we
  surgically collect this image from the cerebellum of a mouse, with
  Purkinje neurons outlined).}
\vspace{-6mm}
\label{fig:purkinje-biology}
\end{minipage}
}
\end{figure}

\vspace{2mm}{\noindent \bf When can we use low power modes?} To use
low power modes until neuronal activity of interest, we must define
the notion of interesting activity. Since neural implants are used for
many tasks, this definition can vary. Our study monitors Purkinje
neurons in the cerebellum. Purkinje firing therefore constitutes our
activity of interest, and can be separated into two classes --
unsynchronized and synchronized firing. To understand these firing
patterns, consider Figure \ref{fig:purkinje-biology}(a), which shows
the cellular anatomy of a Purkinje neuron.

Cerebellar Purkinje neurons are driven by two inputs. The first is a
set of parallel fibers which relay activity from other parts of the
cerebellum. Parallel fibers are connected to Purkinje neurons using
the spindly outgrowths of the neurons, i.e., dendrites.  The second
input is the inferior olivary nucleus, which provides information
about sensorimotor stimulation \cite{zeeuw:microcircuitry}. Inferior
olivary nuclei are connected to climbing fibers, which in turn feed
Purkinje dendrites.

When either the parallel fibers or the inferior olive fire, spikes are
activated on the Purkinje neuron. These spikes drive the deep
cerebellar nucleus, influencing motor control and longer-term
cerebellar plasticity \cite{ozden:reliable}. The exact nature of
Purkinje activity depends on the triggering input. Purkinje spikes due
to parallel fibers occur at 17-150 Hz, while those prompted by the
inferior olivary nuclei occur at 1-5 Hz \cite{ozden:reliable}.

Neuroscientists are studying many aspects of Purkinje spiking, but one
that is important is that of synchronized spiking
\cite{ozden:reliable, tank:spatially, sullivan:invivo,
  kaloshi:visual}. While single Purkinje neurons usually fire
seemingly in isolation, occasionally clusters of Purkinje neurons fire
close together in time. Such synchronized firing usually occurs when
neighboring olivary nuclei are activated in unison. Figure
\ref{fig:purkinje-biology}(b) shows imaging data we collect from an
anesthetized mouse, where Purkinje neurons have been outlined. The
flashing neurons represent firing while those in black represent
quiescence. In the time slice shown, several Purkinje neurons fire
synchronously.

Given their importance, synchronized firing is our focus. We enable
energy-efficiency by using low power modes when Purkinje
synchronization is absent, and using nominal operation when
synchronized activity occurs. In so doing, we sustain longer battery
life to collect longer and more thorough neuronal recording data for
brain mapping studies.

\vspace{2mm}{\noindent \bf Why do we need neuronal activity
  prediction?} One may initially expect to achieve energy efficiency
by placing the processor in sleep mode until the microelectrode array
captures synchronized Purkinje activity. At this point, the processor
could be transitioned to nominal operating frequency. The problem with
this approach is that scientists are curious not just about
synchronized activity, but also about milliseconds of individual
neuronal activity leading up to synchronized firing
\cite{ozden:reliable, sullivan:invivo}. Hence, a better approach is to
anticipate synchronized activity sufficiently ahead of time so that
events leading up to it are also recorded as often as possible. This
necessitates neuronal prediction.

Several prediction strategies initially spring to mind. For example,
one could try detecting olivary nuclei activity as an accurate
predictor of Purkinje activity. Unfortunately, detecting this activity
and separating its effects from those of the parallel fibers requires
complex analysis that consumes valuable resources and energy in
itself. Alternately, if synchronized firing were to occur with
well-defined periodicity, prediction would be simple, but this is not
the case \cite{ozden:reliable, gauck:control}.

We were intrigued by the prospect of using hardware predictors already
on chips today, to also predict neuronal activity. The branch
predictor, in particular, springs to mind, as the binary nature of
taken/not-taken predictions and outcomes maps naturally to the notion
of Purkinje neurons firing or remaining quiet at an instant in
time. Furthermore, modern branch predictors can leverage correlations
among multiple branches -- like correlated branches, the Purkinje
neurons that synchronize once often synchronize repeatedly
\cite{ozden:reliable}. Purkinje neuron correlations thus fit well with
branch predictors, particularly those that capture correlations among
long histories of branches, like perceptron branch predictors.

\vspace{2mm}{\noindent \bf How much energy can we potentially save?}
To answer this question, we model a baseline with a 300 MHz ARM Cortex
M7 processor, and run four workloads often used to process neuronal
recordings (see Sec.~\ref{sec:methodology} for details). The workloads
read the neuronal data picked up by the Utah array and process it to
assess whether it represents synchronized activity. When they identify
synchronized activity, they perform additional signal processing on
all neuronal activity (synchronized or unsynchronized) in the next
500ms. They then again read neuronal data to assess when the next
synchronized event occurs. The baseline does not use the Cortex M7's
idle low power modes because the workloads either continuously profile
the neuronal data to detect synchronized activity, or process neuronal
spiking during and right after synchronization. Without the ability to
predict Purkinje synchronization, the processor cannot know when it is
safe to pause execution and use idle low power modes (though in
Sec.~\ref{sec:results} we do study the potential of using active low
power modes like DVFS, if the Cortex M7 were to support them).

\begin{figure}[t]
\centering
{
\begin{minipage}[t]{0.48\textwidth}
\centering
\vspace{-4mm}
\epsfig{file=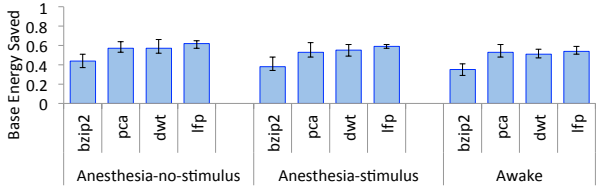, scale=0.36, angle=90}
\vspace{-52mm}
\caption{\small Energy savings due to perfect Purkinje synchronization
  prediction, assuming that four neurons must fire for synchronization
  activity. {\sf Anesthesia-no-stimulus} and {\sf Anesthesia-stimulus}
  represent cases where the mouse is under anesthesia with no stimulus
  and with stimulus respectively, while {\sf Awake} corresponds to
  non-anesthetized mice.}
\vspace{-6mm}
\label{fig:baseline-energy-savings}
\end{minipage}
}
\end{figure}

We contrast the baseline against an ideal -- and hence unrealizable --
oracle neuronal activity predictor that knows the future, and imposes
no area, performance, or energy overheads. This oracle predictor views
time in epochs, and is invoked at the end of each epoch to predict
whether synchronized Purkinje activity is to occur in the next
epoch. Based on the timescales that Purkinje spikes are sustained
\cite{ozden:reliable}, we assume 10ms epochs. We also consider how
many neurons must fire close together in time to be considered
synchronized. While scientists may use different counts depending on
what exactly they are studying \cite{ozden:reliable}, we assume that
four firing neurons constitute a synchronized event for now\footnote{
  Sec.  \ref{sec:results} shows the impact of varying the number of
  neurons that must fire to constitute a synchronized
  event.}. Overall, the processor is suspended in sleep state until
the oracle predictor anticipates synchronization. In response, the
processor transitions to nominal operation, capturing both the 10ms
leadup time to synchronized activity, and the following 500ms of
activity. We also model transition times among low power modes. Since
these take tens of $\mu$s on Cortex M processors, they have little
impact on ms-ranging epochs \cite{arm:transitiontimes}. Figure
\ref{fig:baseline-energy-savings} quantifies the results for three
types of neuronal activity, totaling 26 minutes of Purkinje spiking:

\vspace{2mm}{\noindent \textcircled{1} Anesthesia without stimulus:}
We place mice under deep anesthesia and extract seven 2-minute traces
of 32 Purkinje neurons.  Anesthetized mice exhibit little movement
aside from occasional spontaneous twitching of the limbs, whi\-skers,
and tail. Like past work on synchronized Purkinje firing
\cite{ozden:reliable}, much of the neuronal activity we study focuses
on these experiments since they are the easiest to collect.

\vspace{2mm}{\noindent \textcircled{2} Anesthesia with stimulus:} To
study the effect of controlled sensory stimuli on Purkinje neurons,
like past work \cite{ozden:reliable}, we apply 20-40 psi air puffs
every 40ms to the whiskers of the anesthetized mice. We collect three
traces of Purkinje activity, each 2 minutes long. The sensorimotor
stimulation from the air puffs increases Purkinje synchronization
\cite{ozden:reliable}.

\vspace{2mm}{\noindent \textcircled{3} Awake:} We collect three
2-minute neuronal traces from an awake free-roaming mouse. The rate of
synchronized Purkinje firing varies depending on how the mouse moves.

\vspace{2mm} Figure \ref{fig:baseline-energy-savings} shows that all
benchmarks stand to enjoy significant energy benefits in every single
case. We separate energy benefits into average numbers for each of the
traces in \textcircled{1}-\textcircled{3}, also showing the minimum
and maximum values with error bars. With an ideal Purkinje
synchronization predictor, energy savings can span 29-65\% of total
processor energy. Naturally, as Purkinje synchronizations become more
frequent (either because mice are stimulated with air puffs or are
awake), energy benefits drop since the processor cannot be placed in
sleep mode for quite as long. Still, even in these cases, 63\% of
energy can be saved with ideal predictors.

\section{Implementation}\label{sec:implementation}

Our goal is to show that branch predictors can be used for neuronal
prediction, but there are many potential ways that they could be
architected for this purpose. We study a design that uses the same
branch predictor to predict branches and neuronal activity as it
sufficently demonstrates the branch predictor's ability to predict
Purkinje activity and guide energy management. However, it may be even
better to implement separate branch/neuronal predictors. Our work sets
the foundation for future studies of these alternate design points.

\subsection{Energy Management Strategies}\label{sec:energy-management-strategies}
\begin{figure}[t]
\centering
{
\begin{minipage}[t]{0.48\textwidth}
\centering
\vspace{-4mm}
\epsfig{file=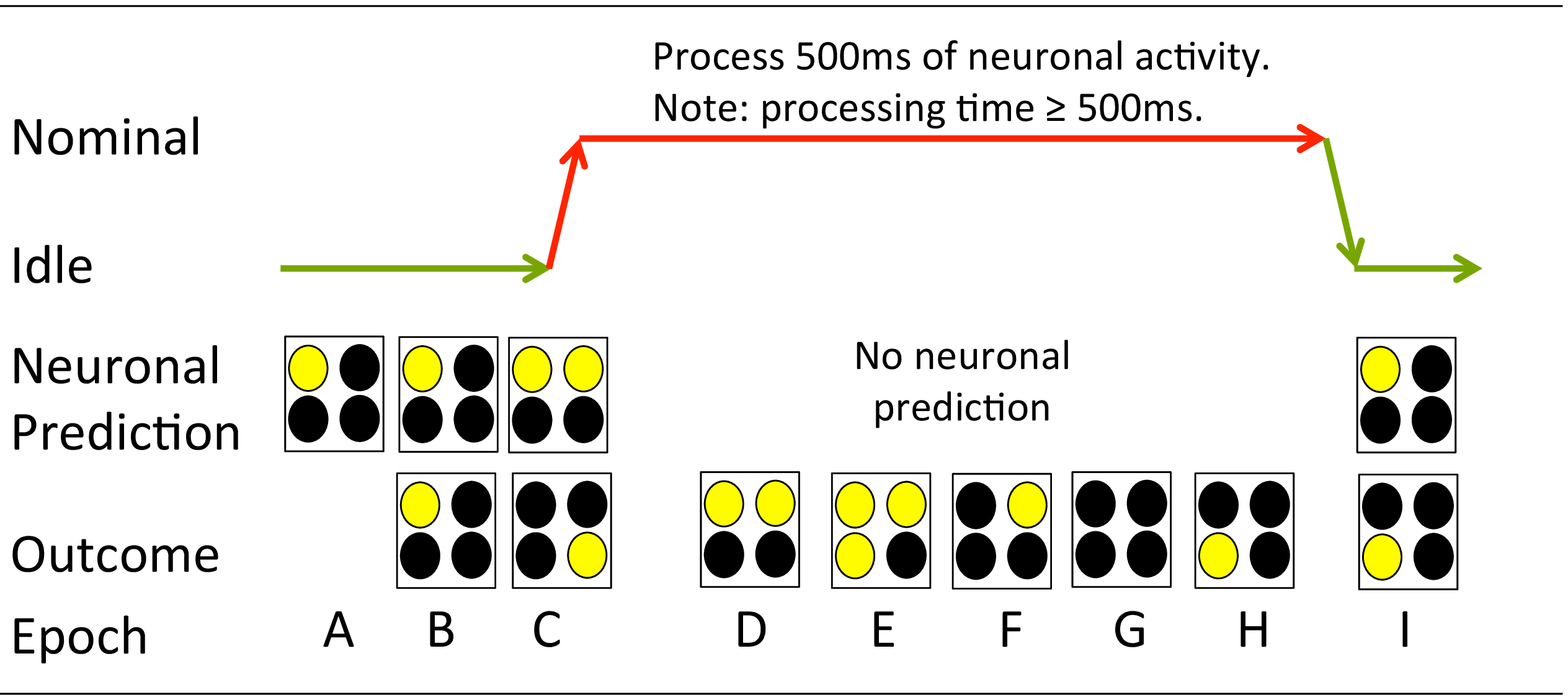, scale=0.32, angle=90}
\vspace{-36mm}
\caption{\small The processor is suspended in idle low power mode, but
  the branch predictor is kept on to predict Purkinje spiking. When it
  correctly predicts synchronized Purkinje firing (in this example,
  two of the four neurons firing), the processor goes to nominal
  operating frequency.}
\label{fig:energy-management-correct-prediction}
\vspace{-6mm}
\end{minipage}
}
\end{figure}

Figure \ref{fig:energy-management-correct-prediction} shows how we
manage energy. Since Purkinje activity is usually unsynchronized, the
Cortex M7 is placed in idle low power mode, turning off processor and
cache clocks but continuing to power DRAM. Our {\sf Idle} state is
different from traditional low power modes in an important way -- we
keep the branch predictor on, and as described in
Sec.~\ref{sec:branch-brain-predictor-implementation}, implement a
hardware FSM to use the predictor to predict Purkinje firing instead
of branches. Figure \ref{fig:energy-management-correct-prediction}
shows {\sf Neuronal Predictions} and actual {\sf Outcomes}, split into
{\sf Epochs} of time labeled {\sf A-I} (which are 10ms in our
studies). Our example monitors four neurons, shown in circles. Yellow
circles represent firing, while black circles represent quiescence;
synchronization occurs when at least two neurons fire.

In epoch {\sf A}, the branch predictor is used for neuronal prediction
and anticipates only a single Purkinje neuron firing in the next
epoch, {\sf B}. Consequently, the processor continues in {\sf
  Idle}. The prediction is correct as it matches the {\sf Outcome} in
{\sf B}\footnote{Sec. \ref{sec:branch-brain-predictor-implementation}
  explains that neuronal activity outcomes are provided by the Utah
  array and ADCs, which place data in DRAM.}. Simultaneously in {\sf
  B}, the predictor predicts that only one neuron will fire in {\sf
  C}. This turns out to be correct again -- although the exact neuron
that fires does not match the prediction, a concept that we will
revisit shortly -- and the processor continues in {\sf Idle}. However,
in {\sf C}, the predictor anticipates a synchronization event between
the two top Purkinje neurons. Consequently, the processor is
transitioned into {\sf Nominal} operating mode. Since transition times
on the Cortex M7 are orders of magnitude smaller than 10ms epoch times
\cite{arm:transitiontimes}, our prediction enables us to awaken the
processor sufficiently early to process not only synchronization
activity but also activity leading up to it. Once in nominal
operation, the processor analyzes 500ms of Purkinje neuron activity,
which can consist of synchronized and unsynchronized spiking, as shown
in {\sf D-E} and {\sf F-H} respectively. During this time, the branch
predictor returns to predicting branches and not neuronal
activity. Note that the time taken to analyze 500ms of neuronal
activity can exceed 500ms. Finally, the processor again transitions to
{\sf Idle}, with the branch predictor returning to brain activity
prediction. Overall, there are four possible combinations of neuronal
prediction and outcomes:

\vspace{2mm}{\noindent \bf \textcircled{1} Correctly predicted
  non-synchronization:} This is desirable to idle the processor as
long and often as possible.

\vspace{2mm}{\noindent \bf \textcircled{2} Correctly predicted
  synchronization:} We want most synchronizations to be correctly
predicted, enabling capture of both the activity before
synchronization, as well as 500ms of neuronal activity during and
after it.

\begin{figure}[t]
\centering
{
\begin{minipage}[t]{0.48\textwidth}
\centering
\vspace{-4mm}
\epsfig{file=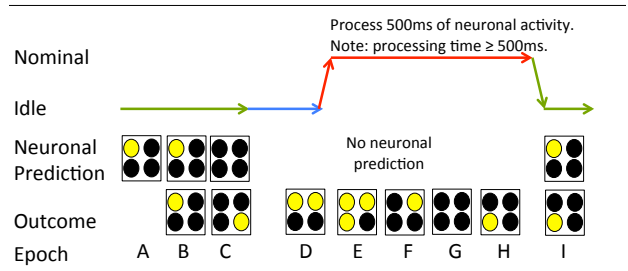, scale=0.32, angle=90}
\vspace{-36mm}
\caption{\small The branch predictor can also mispredict neuronal
  activity. In this figure, it misses upcoming Purkinje
  synchronization, so the processor does not record 10ms events leading
  up to synchronization (in blue), though it is woken up when the
  misprediction is identified.}
\vspace{-4mm}
\label{fig:energy-management-wrong-prediction}
\end{minipage}
}
\end{figure}

\vspace{2mm}{\noindent \bf \textcircled{3} Incorrectly predicted
  non-synchronization:} Occasionally, branch predictors may make
mistakes. Figure \ref{fig:energy-management-wrong-prediction} shows
that in epoch {\sf C}, the branch predictor expects no Purkinje
neurons to fire in epoch {\sf D}. Unfortunately, this prediction turns
out to be incorrect, as the top two Purkinje neurons do fire in {\sf
  D}. We mitigate the damage caused by this by transitioning to {\sf
  Nominal} operating mode as soon as we detect the
misprediction. Unfortunately though, the implant still misses the
opportunity to monitor pre-synchronization activity. Therefore, we aim
to reduce the incidence of this type of misprediction. Note that
technically, this kind of misprediction actually saves {\it more}
energy because it runs the processor in {\sf Idle} for longer (see the
blue arrow in Figure
\ref{fig:energy-management-wrong-prediction}). However, since it
misses important pre-synchronization activity, this type of energy
saving is actually undesirable. Overall, we use low power modes to
save energy, running the risk that we occasionally mispredict neuronal
activity and lose some pre-synchronization activity. But if we reduce
the incidence of this type of misprediction, this tradeoff is
worthwhile since we ultimately sustain far longer battery life and
collect considerably longer neuronal activity recordings overall.

\vspace{2mm}{\noindent \bf \textcircled{4} Incorrectly predicted
  synchronization:} Finally, it is also possible for the branch
predictor to incorrectly predict synchronized behavior, only to find
that this behavior does not occur in the next epoch. This represents
wasted energy usage as the processor is transitioned to {\sf Nominal}
operation unnecessarily. However, as soon as we detect no Purkinje
synchronization in the following epoch, we transition the processor
back to {\sf Idle} mode.

\vspace{2mm} Recall that in Figure
\ref{fig:energy-management-correct-prediction}, the branch predictor
predicted, in {\sf B}, that the upper left neuron would fire in {\sf
  C}. Ultimately the lower right neuron fired. We refer to such
predictions as {\it accidentally correct} as they represent situations
where higher-level prediction of synchronization is correct even
though the prediction of the individual Purkinje neurons are
wrong. Accidentally correct predictions can occur in many ways -- for
example, the predictor may anticipate a synchronized event with the
top two neurons firing, even though the bottom two neurons ultimately
fire. While accidentally correct predictions enable correct operation,
our goal is to design predictors that are correct in a robust manner,
and do not rely on ``accidental luck''. We therefore focus on accuracy
for both per-neuron and synchronized prediction.

Finally, our branch predictor loses state when switching between
standard branch prediction and neuronal prediction. This is similar to
the impact of context switches \cite{jimenez:perceptron}. We have
found that compared to a baseline that does not use idle low power
modes, our approach prompts a 4\% drop in branch average prediction
accuracy during nominal operation. Since this does not have any
discernible performance effect on our implant, and because energy
savings are immense (see Sec. \ref{sec:methodology}), we leave more
advanced approaches that save branch predictor state for future work.

\begin{figure}[t]
\centering
{
\begin{minipage}[t]{0.48\textwidth}
\centering
\vspace{-4mm}
\epsfig{file=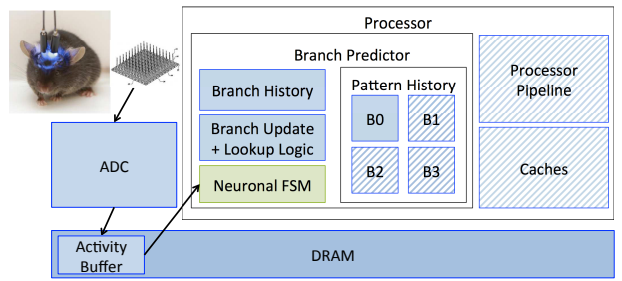, scale=0.32, angle=90}
\vspace{-34mm}
\caption{\small In idle low power mode, striped components are powered
  off, while a hardware FSM co-opts (part of) the conventional branch
  predictor for neuronal prediction. Components are not drawn to
  scale.}
\vspace{-4mm}
\label{fig:hw-overview}
\end{minipage}
}
\end{figure}

\subsection{Branch/Brain Predictor Implementation}\label{sec:branch-brain-predictor-implementation}

Our modifications leave branch predictor access latencies, energy
consumption, etc., {\it unchanged} in normal operating
mode. Therefore, this section focuses on neuronal prediction when the
processor is in low power mode. Figure \ref{fig:hw-overview} presents
the hardware required by our proposal. On the left, we show a mouse
with an embedded implant and a Utah microelectrode array used to read
out Purkinje activity. This activity is amplified and digitized by the
ADC, and stored in a designated DRAM location called an {\sf activity
  buffer}.

An important first step in processing the Utah array's data is to
identify which conductive silicon needles on the array correspond to
Purkinje neurons. Recall that the Utah array has hundreds of
needles. Many of these needles probe non-neuronal tissue, while others
probe neurons. Many implants therefore run signal processing code at
installation time to associate needles to specific neurons by studying
1-2 seconds of neuronal activity \cite{kwon:neuroquest}. Since the
implant stays in place, once calibration completes, we know exactly
which of the Utah array needles correspond to Purkinje neurons. The
implant is then free to run its workloads.


Figure \ref{fig:hw-overview} shows that when these workloads are
placed in low power mode, the pipeline and caches are gated off
(indicated by stripes). However, the branch predictor is treated
differently. We show a general branch predictor structure made up of
pattern history tables and branch history tables\footnote{While our
  example shows one branch history register, the same approach could
  be applied to branch history tables too.}. These branch predictor
structures are looked up and updated using combinational logic, also
shown in the diagram.

When the processor is in idle low power mode, the branch predictor is
used to perform neuronal prediction. One option is to leave the entire
branch predictor structure on for this purpose. However, this is
needlessly wasteful since modern branch predictor tables tend to use
tens of KBs with thousands of entries. Meanwhile, modern recording
technologies allow us to probe the activity of hundreds of neurons
simultaneously \cite{stevenson:how} so we only technically require
hundreds of entries in the branch predictor to make per-Purkinje spike
predictions. Therefore, we exploit the fact that modern branch
predictors are usually banked \cite{baniasadi:branch, parikh:power}
and turn off all but one bank. This bank suffices to perform neuronal
prediction. To enable this bank to remain on while the remainder of
the branch predictor is power gated, we create a separate power domain
for it. This requires a separate set of high Vt transistors and
control paths. We model the area, timing, and energy impact of these
changes (see Sec. \ref{sec:methodology}).

Figure \ref{fig:hw-overview} shows that we add a small {\sf neuronal
  FSM} (shown in green). We modify the code run in the calibration
step after implant installation to add a single store instruction. This
updates the contents of a register in the {\sf neuronal FSM}
maintaining a bit-vector used to identify which of the Utah array's
silicon needles probe Purkinje neurons. The {\sf neuronal FSM} uses
this bit vector to decide which entries in the {\sf activity buffer}
store activity from neuronal (rather than non-neuronal) tissue. The
{\sf neuronal FSM} then co-opts the branch predictor for neuronal
prediction during idle low power mode. It orchestrates two types of
operations, every time epoch:

\begin{figure}[t]
\centering
{
\begin{minipage}[t]{0.48\textwidth}
\centering
\vspace{-4mm}
\epsfig{file=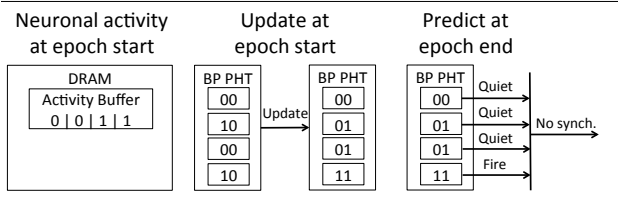, scale=0.32, angle=90}
\vspace{-44mm}
\caption{\small The branch predictor predicts neuronal activity in
  each epoch in low power mode. We show the Smith predictor as an
  example.}
\vspace{-6mm}
\label{fig:update-predict}
\end{minipage}
}
\end{figure}

\vspace{2mm}{\noindent \textcircled{1} Updates with neuronal
  outcomes:} In every epoch, we first update predictor tables with
recent Purkinje activity. Consider Figure \ref{fig:update-predict},
which shows the DRAM {\sf activity buffer} at the start of the
epoch. The activity buffer maintains an entry for every conductive
needle in the Utah array indicating firing (a bit value of 1) and
quiescence (a bit value of 0). The {\sf neuronal FSM} is programmed to
know which of these bits correspond to actual neurons via the
calibration step. Our example shows entries for four conductive
needles probing four neurons, two of which remained quiet and two of
which fired at the start of the epoch. Consequently, the {\sf neuronal
  FSM} updates the branch predictor bank left on by treating each
neuron as a separate branch and updating in a manner that mirrors
conventional branch predictor updates. Figure \ref{fig:update-predict}
shows this for a Smith branch predictor with 2-bit saturating counters
and hysteresis \cite{lee:branch, smith:study}. The four branch
predictor entries are used for neurons 0-3, and are updated using the
state machine of a standard 2-bit branch predictor. 

\vspace{2mm}{\noindent \textcircled{2} Predictor lookups for neuronal
  predictions:} Figure \ref{fig:update-predict} sho\-ws that at the
end of the epoch, the {\sf neuronal FSM} must predict whether Purkinje
synchronization will occur in the next epoch. Each neuron's branch
predictor entry is looked up to predict whether that neuron will
fire. In our example, the first three neurons are predicted to remain
quiet while the last one is predicted to fire. Combinational logic
assessses whether enough neurons are predicted to fire to constitute
synchronization. For our example in
Sec. \ref{sec:energy-management-strategies}, where at least two
neurons must fire for synchronization, the {\sf neuronal FSM} assesses
that the next epoch will not see synchronization and hence the
processor can continue in idle low power mode.

\vspace{2mm} While we do not change access times for branch
prediction, we consider timing when the branch predictor is used for
neuronal prediction. Using detailed circuit modeling, we have found
that since neuronal spiking times (and epoch times) range in the order
of milliseconds, the timing constraints of modern branch predictors,
which are typically designed for hundreds of MHz and GHz clocks, are
easily met.

Finally, an important question is why neuronal prediction is needed in
the first place. One may consider sizing the DRAM {\sf activity
  buffer} to be sufficiently large to store lead-up activity to
synchronization. The processor could be transitioned to nominal
operation when sychronization occurs. This approach seemingly
preserves lead-up activity to synchronization without needing
synchronization prediction. Unfortunately, while this approach
suffices for some neuronal implants, it is not a solution for implants
which must process neuronal activity as soon as it occurs, rather than
deferring to a later time. For example, implants designed to provide
stimuli to the brain immediately when a specific neuron fires
\cite{zanos:neurochip2, open-ephys} cannot adopt this approach. Our
goal to enable energy management on {\it all} implant types --
therefore, we tackle the harder problem of neuronal prediction.

\vspace{-2mm}
\section{Branch and Brain Predictors}\label{sec:branch-brain-prediction}


\subsection{Prediction Strategies}\label{sec:prediction-strategies}
A comprehensive study of all branch predictors is beyond the scope of this
paper. Instead, we provide intuition on how their design principles
need to be rethought for neuronal prediction. To accomplish this, we
focus on:

\vspace{2mm}{\noindent \bf Smith predictors:} These use 2-bit
saturating counters with hysteresis (see Figure
\ref{fig:update-predict}). Each Purkinje neuron is allotted a counter
in the prediction table\footnote{Since modern predictors maintain
  large tables with far more entries than the tens-hundreds of neurons
  we can currently record, we assume one entry per neuron and no
  aliasing.}. A branch/neuron's local history is used to predict
future behavior (i.e., correlations among branches/neurons are not
exploited). We have found that local history can, to some extent,
enable prediction of future activity. But, this approach is too simple
to perform accurate neuronal prediction when the mouse roams around
and hence sees more complex cerebellar spiking. 

\begin{figure}[t]
\centering
{
\begin{minipage}[t]{0.48\textwidth}
\centering
\vspace{-4mm}
\epsfig{file=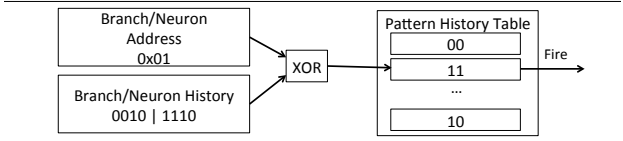, scale=0.32, angle=90}
\vspace{-52mm}
\caption{\small Adapting gshare predictors for neuronal prediction.}
\vspace{-4mm}
\label{fig:gshare}
\end{minipage}
}
\end{figure}

\vspace{2mm}{\noindent \bf Gshare predictors:} Smith predictors do not
exploit correlations between branches, and hence neurons. This is a
problem for Purkinje neurons, which form ``micro-bands'' or grou\-ps
where neurons fire close together in time in a synchronized manner
\cite{ozden:reliable, sullivan:invivo}. To exploit micro-bands, we
study branch predictors that exploit correlated branches. Gshare is a
well-known example of such a predictor. Figure \ref{fig:gshare} shows
how gshare predictors can be co-opted for neuronal prediction. The
{\sf neuronal FSM} from Sec. \ref{sec:implementation} looks up the
predictor table for each individual neuron. If enough of them are
predicted to fire, a synchronization event is predicted. Figure
\ref{fig:gshare} illustrates lookup for neuron number 1. Gshare
performs an exclusive-or between this ``address'' (or neuron number)
and an {\sf n}-bit global history register, which records the last
{\sf n} branch outcomes globally. For neuronal prediction, one could
similarly record spiking behavior of the last {\sf n} neurons. There
are two options -- (a) we can record whether there was Purkinje
synchronization in the last {\sf n} epochs (one bit per epoch); or (b)
whether each of {\sf j} individual neurons in the last {\sf k} epochs
(where {\sf n} equals {\sf j}$\times${\sf k}) fired or not. Recall
that our goal is to perform accurate per-neuron predictions, not just
synchronization predictions (see
Sec. \ref{sec:branch-brain-predictor-implementation}).  We therefore
do (b). Figure \ref{fig:gshare} shows a global history register that
stores activity from four neurons in the last two epochs.

\begin{figure}[t]
\centering
{
\begin{minipage}[t]{0.48\textwidth}
\centering
\vspace{-4mm}
\epsfig{file=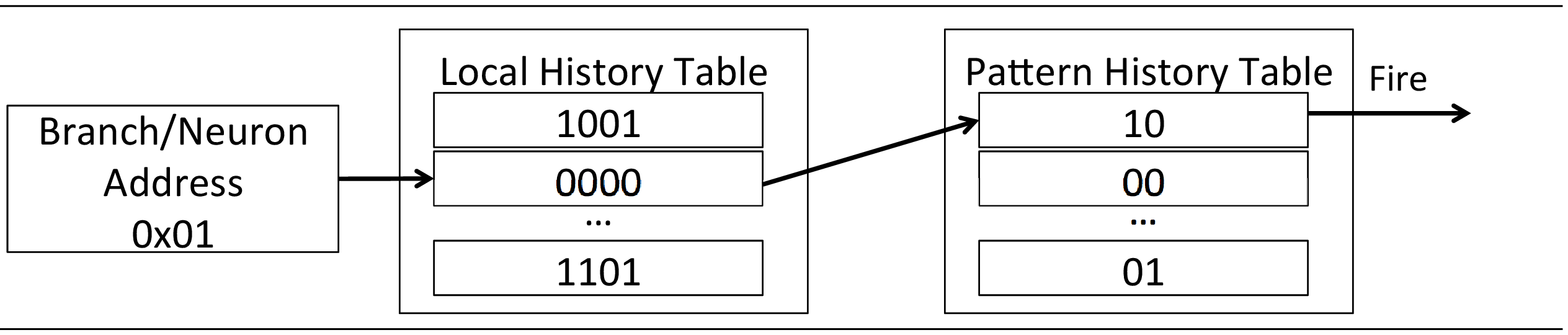, scale=0.32, angle=90}
\vspace{-54mm}
\caption{\small Adapting two-level predictors for neuronal
  prediction.}
\vspace{-6mm}
\label{fig:two-level-adaptive}
\end{minipage}
}
\end{figure}

\vspace{2mm}{\noindent \bf Two-level adaptive predictors:} It is also
possible to exploit inter-branch/neuronal correlations by not just
maintaining global history but also per-branch histories. Two-level
adaptive predictors leverage this insight \cite{yeh:two}. Figure
\ref{fig:two-level-adaptive} co-opts this approach for neuronal
prediction, focusing on the lookup for neuron 1. The neuron number,
like the branch address, indexes a table of local history
registers. The {\sf n}-bit registers record outcomes of the last {\sf
  n} branches and neuronal data that map to that location. In Figure
\ref{fig:two-level-adaptive}, the local history tables store
information on how neuron 1 spiked in the last four epochs (in our
example, neuron 1 was quiet in all four epochs). This history selects
a saturating counter from the pattern history table, which ultimately
determines a prediction for neuron 1 in the next epoch.

\begin{figure}[t]
\centering {
\begin{minipage}[t]{0.48\textwidth}
\centering
\vspace{-4mm}
\epsfig{file=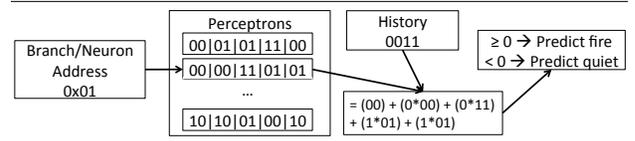, scale=0.32, angle=90}
\vspace{-52mm}
\caption{\small Adapting perceptron predictors for neuronal prediction.}
\vspace{-6mm}
\label{fig:perceptron}
\end{minipage}
}
\end{figure}

\vspace{2mm}{\noindent \bf Perceptron predictors:} Perceptron-based
approaches, inspired by machine learning techniques, are best able to
leverage inter-branch/neuron correlations. Figure \ref{fig:perceptron}
illustrates the operation of a conventional perceptron predictor and
how we can adapt it for neuronal prediction
\cite{jimenez:perceptron}. A table of perceptrons -- rather than
saturating counters -- is looked up for each branch or neuron for
which we desire prediction. Each perceptron entry maintains weights
for each correlated branch/neuron. Like conventional perceptron
prediction for branches \cite{jimenez:perceptron, jimenez:fast}, we
use the weights and prior branch outcomes, stored in the history
register, to calculate:

{\begin{center}
 $y = w_{0} +
\sum\limits_{i=1}^{n}x_{i}w_{i}$ 
\end{center}
}

\noindent Here, $w_{0}$ is a bias weight, $w_{i}$ are weights for
correlated branc\-hes/neurons, $x_{i}$ are prior branch/neuron
outcomes, and $y$ is the output prediction. If $y$ is non-zero, the
branch/neuron is predicted taken/fired. In Figure
\ref{fig:perceptron}, we use 2-bit weights though actual
implementations use 8-bit integer weights \cite{jimenez:perceptron,
  jimenez:piecewise, jimenez:fast}. The weights record a
branch/neuron's dependence on its past behavior through a bias weight,
and its dependence on other (four other, in our example)
branches/neurons through other weights. All values are stored in one's
complement, like the original design \cite{jimenez:perceptron}, with
large positive and negative values indicating positive and negative
correlations respectively. Figure \ref{fig:perceptron} shows that the
looked-up neuron is weakly correlated with its past (a bias weight of
00) but is positively correlated with neurons 2 and 3 (weights of 01),
and strongly but negatively correlated with neuron 1 (weight of 11).

During neuronal prediction, we perform updates in a manner similar to
branch prediction \cite{jimenez:perceptron}. When the {\sf neuronal
  FSM} reads the Purkinje spiking outcomes, it checks if there was a
misprediction or if the weighted sum's magnitude was less than a
threshold $\theta$ (to gauge if training is complete). For either of
these situations, the perceptron entry is updated. The algorithm
increments the $i$th weight if the branch/neuron outcome agrees with
$x_{i}$ and decrements it otherwise. We assume the $\theta$ values
used in prior work for branches \cite{jimenez:perceptron}; we have
found them to suffice for neuronal prediction too.

Because the size of perceptrons scales linearly with the number of
correlated branches/neurons, perceptrons exploit longer branch/neuron
correlation histories than other schemes, which scale
exponentially. This makes perceptron preditions effective at capturing
Purkinje micro-bands. Furthermore, the two problems typically
associated with perceptron predictors -- i.e., high access latency and
power consumption \cite{jimenez:piecewise, amant:high} -- are less of
a concern in our design. Access latencies are usually a problem on
high-performance GHz-range processors with tight cycle times. Instead,
in our system, the processor requires branch predictions on a 300MHz
clock or neuronal prediction in 10ms epochs, which perceptron
predictors using Wallace tree adders \cite{jimenez:neural} can
comfortably achieve. And despite the higher energy requirements of
perceptron predictors, their ability to accurately predict neuronal
activity enables more aggressive use of low power mod\-es and hence
much lower overall system energy.

Figure \ref{fig:perceptron} focused on perceptron prediction using
global branch/neuron history. Naturally, it is also possible to
perform two-level prediction with perceptrons, where a table of
per-branch/neuron histories is first indexed, before looking up the
perceptrons. We study this approach too.

\subsection{Lessons Learned}\label{sec:lessons-learned}

We summarize key lessons (see Sec. \ref{sec:results} for details).

\vspace{2mm}{\noindent \bf \textcircled{1} Correlations matter more
  than local history:} A neuron's history can provide some indication
of future behavior. But local history must be coupled with
inter-neuronal correlations for good prediction accuracy. This is
because micro-bands of correlated neurons synchronize
\cite{ozden:reliable}, and predictors that can exploit longer
histories of correlated neurons are hence most
accurate. Area-equivalent perceptron predictors can achieve 35\%
prediction accuracy over Smith predictors. Perceptrons also outperform
gshare and adaptive predictors.

\vspace{2mm}{\noindent \bf \textcircled{2} Correlations remain
  important in the presence of sensorimotor stimulation:} When we blow
air puffs on the whis\-kers of anesthetized mice or study free-roaming
mice, Purkinje activity continues to be correlated and synchronization
becomes more frequent. Smith predictors, wh\-ich rely on only local
history, drop off in accuracy. For example, awake mice see an average
of 27\% accuracy, while gshare and two-level adaptive approaches
achieve average accuracies of only 35\%. Perceptrons continue to
exploit inter-neuronal correlations however, and achieve 68\%
accuracy. We also qualitatively observe that when awake mice move
more, perceptron predictors become are more accurate than other
approaches.

\vspace{2mm}{\noindent \bf \textcircled{3} Prediction accuracy trumps
  higher energy needs:} Two-level adaptive and perceptron approaches
consume more po\-wer than simpler Smith predictors. We find, however,
that complex predictors, especially perceptrons, predict neuronal
activity so much more accurately that they can use low power modes
aggressively enough to save energy overall.

\begin{figure}[t]
\centering {
\begin{minipage}[t]{0.48\textwidth}
\centering
\vspace{-6mm}
\epsfig{file=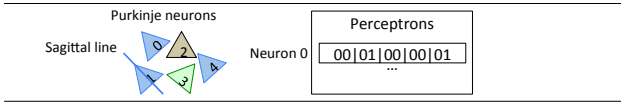, scale=0.32, angle=90}
\vspace{-58mm}
\caption{\small Parasagittal Purkinje neurons spaced a few micron
  apart are usually correlated. This is reflected in their perceptron
  table entries.}
\vspace{-6mm}
\label{fig:parasagittal}
\end{minipage}
}
\end{figure}

\vspace{2mm}{\noindent \bf \textcircled{4} Neurons experience ``phase
  changes'':} Branch mispredictions often occur when branch
correlations change, or when branches are not {\it linearly separable}
for perceptrons. Linear separability refers to the fact that
perceptrons can perfectly predict only branches whose Boolean function
over variables $x_{i}$ have its true instances separated from its
false instances with a hyperplane, for some values of $w_{i}$
\cite{jimenez:neural}. Similarly, there are situations when perceptron
predictors achieve only 30\% neuronal prediction accuracy (and other
predictors achieve even less) because many {\it neurons} are not
linearly separable. This is because neurons, just like branches,
exhibit phase-like behavior. Groups of Purkinje neurons sometimes
switch between micro-bands -- i.e., a neuron changes which other
neurons it correlates with. This well-known biological phenomenon
\cite{ozden:reliable} can lead to mispredictions. We will explore
techniques like piecewise linear branch prediction, which target
linear separability of branches, \cite{jimenez:piecewise} to overcome
this problem for neurons in the future.

\vspace{2mm}{\noindent \bf \textcircled{5} Predictors can capture
  brain physiology:} Parasagittal Purkinje neurons spaced micrometers
apart are known to have the highest correlation in behavior
\cite{ozden:reliable}. Figure \ref{fig:parasagittal} shows that
neurons have a sagittal line that divides their bodies into equal left
and right sides. Parasagittal neurons are those that are parallel to
one anothers' sagittal lines. In our example, neurons 0, 1, and 4 are
parasagittal and correlated. We have found that perceptron branch
predictors accurately capture correlations among parasagittal Purkinje
neurons, maintaining much larger weig\-hts for them. We have found
that on average, the weights for parasagittal neurons are 50\%+ larger
than the weights for other neurons. Figure \ref{fig:parasagittal}
shows an example where the weights for neuron 1 and 4 show positive
correlation in neuron 0's perceptron entry.

\vspace{-2mm}
\section{Methodology}\label{sec:methodology}

\begin{table}[t]
\centering
\begin{tabular}{|c|c|}
\hline
{\small Pipeline} & {\small 6-stage, in-order, forwarding}\\\hline
{\small Issue width} & {\small 2-wide}\\\hline
{\small Instruction and data cache} & {\small 32KB with ECC}\\\hline
{\small Baseline branch predictor} & {\small 8KB Smith predictor}\\\hline
{\small Integer/FPU} & {\small 4-stage/5-stage pipe}\\\hline
{\small Register file} & {\small 6/4 read/write ports}\\\hline
\end{tabular}
\label{tab:simulator}
\vspace{-2mm}
\caption{\small Parameters of our system.}
\vspace{-6mm}
\end{table}

{\noindent \bf Simulation infrastructure:} We model a processor
similar to the ARM Cortex M7, with the configuration of Table 1. This
paper performs early-stage design exploration. Therefore, rather than
implement the chip in hardware, we rely on careful cycle-accurate
software simulation. Our processor runs at 300MHz and maintains the
standard Cortex M7 idle low power mode whe\-re pipelines and caches
can be gated off to save power. We use CACTI
\cite{muralimanohar:cacti} and McPAT \cite{li:mcpat} for power/energy
analysis. We model the area, timing, and energy implications of
creating a separate power domain for the branch predictor bank for
neuronal prediction. The additional Vt transistors and control
wiring/logic increases chip area by 1.4\%. Branch prediction acces
latencies remain unchanged, however. Further, we model the addition of
the {\sf neuronal FSM} as part of our analysis. In general, we find
that its simplicity means that it can be implemented with
area-efficient and energy-efficient combinational logic.

\vspace{2mm}{\noindent \bf Workloads:} We use four neuronal spiking
analysis workloads, selected for their common use in the neuroscience
community, to extract biologically relevant data from neuronal
recordings \cite{kwon:neuroquest}. The four workloads are:

\vspace{2mm}{\noindent \textcircled{1} Compression:} We use
\textbf{\textsf{bzip2}} to compress the spiking data recorded by the
ADC for 500ms after synchronization.

\vspace{2mm}{\noindent \textcircled{2} Artifact removal:}
Microelectrode arrays can often pick up noise caused by muscle
movement in the scalp, jaws, neck, body, etc. These artifacts can be
removed with principal component analysis. Our \textbf{\textsf{pca}}
benchmark stacks the data from our electrodes, and for each electrode,
projects into the PCA domain, yielding cleaned signals
\cite{kwon:neuroquest}.

\vspace{2mm}{\noindent \textcircled{3} LFP extraction:} In
\textbf{\textsf{lfp}}, we apply a fifth-order Butterworth filter on
the neuronal data to enhance low-frequency signals in the range of
0.5-300Hz, as is common practice \cite{kwon:neuroquest}.

\vspace{2mm}{\noindent \textcircled{4} Denoising:} Reducing noise in
neuronal recordings is an important step in neuronal processing. There
are several ways to denoise, but we use discrete wavelet transforms or
\textbf{\textsf{dwt}} with Rigrsure thresholding, similar to prior work
\cite{kwon:neuroquest}.

\vspace{2mm}{\noindent \bf Mouse surgeries:} To perform early design
space exploration, we extract neuronal activity from mice in vivo,
before actually designing the implant. We rely on the emerging field
of optogenetics for this. Optogenetics gives neuroscientists the
ability to use pulses of light to image and control almost any type of
neuron in any area of the brain, with precise timing. We perform
surgeries on C57BL/6 mice on postnatal days 21-42. We perform small
craniotomies of approximately 2mm diameter over lobule 6 locations on
the mice cerebella, from which we collect Purkinje activity. For mice
under anesthesia, we use ketamine/xylazine to achieve deep ansthetized
state. Further, we load the Purkinje cells of the area of study with
calcium indicator Oregon Green BAPTA-1/AM (Invitrogen), as described
previously \cite{sullivan:invivo}. This indicator flouresces under
exposure to light, allowing us to collect images such as Figure
\ref{fig:purkinje-biology} using two-photon microscopes
\cite{ozden:reliable}. We track the activity of 32 Purkinje neurons.

\vspace{-2mm}
\section{Results}\label{sec:results}
Predictor organizations (i.e., pattern history table and bran\-ch
history register dimensions, etc.) that maximize neuronal prediction
may be different from those that optimize traditional branch
prediction. But since branch prediction has been studied for decades,
our focus is on neuronal prediction in this paper. We will study the
best ``compromise'' branch/neuronal predictor organizations in future
work.

\begin{figure}[t]
\centering
{
\begin{minipage}[t]{0.48\textwidth}
\centering
\vspace{-4mm}
\epsfig{file=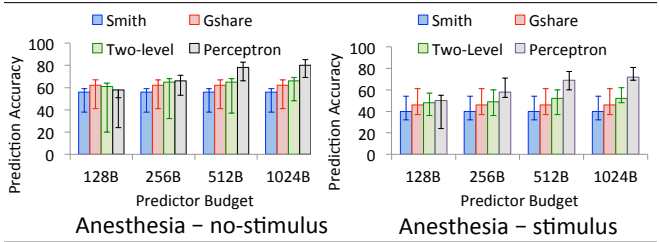, scale=0.36, angle=90}
\vspace{-48mm}
\caption{\small Prediction accuracies for mice under anesthesia
  without stimulus, and with air puffs blown into their whisker
  pads. Prediction accuracies are shown as a function of the hardware
  budget available for the bank of the branch predictor left open.}
\vspace{-4mm}
\label{fig:anesthesia-accuracy}
\end{minipage}
}
\end{figure}

\begin{figure}[t]
\centering
{
\begin{minipage}[t]{0.48\textwidth}
\centering
\vspace{-4mm}
\epsfig{file=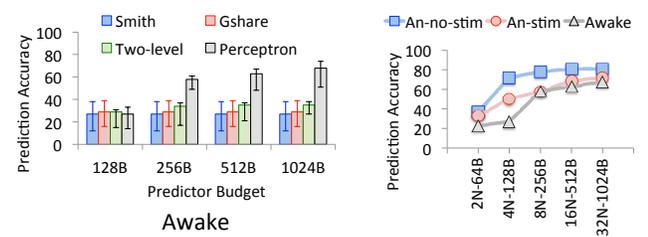, scale=0.36, angle=90}
\vspace{-46mm}
\caption{\small (Left) Prediction accuracy for awake free-roaming
  mice, as a function of the predictor area budget; and (right)
  perceptron predictor accuracy as a function of the neuron history
  length.}
\vspace{-4mm}
\label{fig:awake-perceptron-history}
\end{minipage}
}
\end{figure}

\vspace{2mm}{\noindent \bf Per-neuron prediction accuracy:} We first
quantify prediction accuracy for individual Purkinje neurons. We study
the cross product of recordings on awake and anesthetized mice (with
and without air puffs) and our four benchmarks. This represents
13$\times$4 or 52 experiments. We separate results for {\sf Awake} and
anesthetized mice with air puffs ({\sf Anesthesia-stimulus}) and
without ({\sf Anesthesia-no-stimulus}).

Figure \ref{fig:anesthesia-accuracy} presents results for anesthetized
mice. The y-axis plots per-neuron prediction accuracy, with higher
numbers being better. The x-axis shows the hardware budget for the
predictor (one branch predictor bank) from 128 bytes to 1KB. Modern
branch predictors are 8-16KB, and even 1KB represents reasonable size
estimates for bank (which is all we need for neuronal prediction). For
each hardware buget, we have exhaustively studied predictor
organizations and report results from the organization with the best
average accuracy. At each hardware budget, we find that gshare and
two-level predictors perform best when they maintain history for
0.5-0.6$\times$ the neurons as the perceptron.

Figure \ref{fig:anesthesia-accuracy} shows that perceptrons predict
Purkinje neuron activity more accurately than other approaches,
particularly with larger budgets. Smith predictor accuracies remain
flat since even the smallest size (128 bytes) has enough area to
maintain per-neuron counters. Therefore, when hardware budgets
increase to 512 bytes or 1KB, Smith predictors cannot exploit the
additional area, while perceptron predictors can. At larger areas,
predictors that exploit inter-neuron/branch correlation like gshare
and two-level adaptive schemes perform much better. At modest hardware
budgets of 1KB, perceptron predictors achieve average prediction
accuracy of 80\%, and as high as 85\%. Perceptrons become even more
accurate compared to other approaches when sensorimotor stimulation --
and hence the complexity of Purkinje activity -- increases (see {\sf
  Anesthesia-stimulus} results). While blowing air puffs into
whis\-kers does make it difficult to predict neuronal behavior,
perceptron branch predictors still achieve accuracies of 72\% for 1KB
hardware budgets.

The left side of Figure \ref{fig:awake-perceptron-history} shows
results for awake mice. The increased complexity of neuronal spiking
pro\-mpts Smith, gshare, and two-level adaptive predictors to
mispredict more often. The two-level adaptive scheme only achieves
35\% accuracy. However, perceptrons still achieve an average of
68\%. Qualitatively, we found that prediction accuracy varies more
dramatically when the mouse moves not only its tail but also its limbs
(see the larger min/max bars).


The graph on the right of Figure \ref{fig:awake-perceptron-history}
shows how perceptrons achieve more accuracy. We show accuracy as we
vary the number of weig\-hts stored in each perceptron entry. A label
of {\sf jN-kB} on the x-axis indicates 8-bit integer weights for {\sf
  j} potentially correlated neurons, totaling {\sf k} bytes (assuming
that we need a separate perceptron entry for every neuron we want to
predict). The larger {\sf k} is, the greater the correlation history
amongst branches/neu\-rons, and the more accurate our neuronal
prediction. When we plotted this data, we noticed an interesting
relationship between the number of weights required for accurate
predictions and the biology of Purkinje neurons. Studies have shown
that usually a handful (2 to 8) of neurons form micro-bands
\cite{ozden:reliable}. The graph on the right of Figure
\ref{fig:awake-perceptron-history} mirrors this observation, with
sharp accuracy benefits when the number of weights in the perceptron
goes from 2 to 8, particularly when the mouse is awake.

\begin{figure}[t]
\centering
{
\begin{minipage}[t]{0.48\textwidth}
\centering
\vspace{-4mm}
\epsfig{file=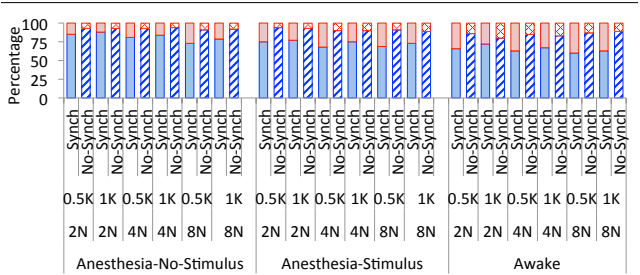, scale=0.36, angle=90}
\vspace{-42mm}
\caption{\small Percentage of synchronized events predicted correctly
  (solid blue) and incorrectly (solid red), and percentage of
  unsynchronized events predicted correctly (striped blue) and
  incorrectly (striped red). We show average results as we vary the
  number of neurons in a synchronized event from 2, 4, to 8, and as
  the predictor size varies between 512 bytes and 1KB. All results are
  for perceptron predictors.}
\vspace{-6mm}
\label{fig:synch-prediction-accuracy}
\end{minipage}
}
\end{figure}

\vspace{2mm}{\noindent \bf Synchronization prediction accuracy:} So
far, we have discussed prediction accuracy for each individual
Purkinje neuron's behavior. However, our aim is to ultimately predict
synchronization behavior. We focus on the perceptron predictor for
these studies as they are far more accurate than other
approaches. While good prediction accuracy for individual neurons is a
good indicator of synchronization prediction, their relationship is
complicated by two competing factors. On the one hand, accidentally
correct predictions may occur (see
Sec. \ref{sec:energy-management-strategies}), boosting synchronization
prediction accuracy. On the other hand, synchronization requires {\it
  multiple} neurons to be simultaneously predicted correctly. The
probability that multiple neurons are concurrently predicted
accurately is lower than accuracy for an individual neuron.

Figure \ref{fig:synch-prediction-accuracy} summarizes synchronization
prediction accuracy. We separate results for awake and anesthetized
mice, varying the perceptron predictor hardware budget between 512
bytes (0.5KB) and 1KB. We vary the number of neurons that must
simultaneously fire to be considered a synchronized event from 2 to 8
(represented by {\sf 2N}, {\sf 4N}, and {\sf 8N}). For each of these
cases, we plot two bars. The first bar stacks the percentage of total
synchronized events that are correctly predicted (solid blue) and
incorrectly predicted (solid red). The second bar stacks the
percentage of total non-synchronized events that are correctly
predicted (striped blue) and incorrectly predicted (striped red). For
both bars, we desire higher contributions from the blue stacks.

Figure \ref{fig:synch-prediction-accuracy} shows that perceptrons
accurately predict most synchronized and non-synchronized
events. Accuracy increases with larger predictors, but remains
consistently 75\%+ under anesthesia with no stimulus. Naturally,
stimuli and awake states make prediction harder, but perceptrons still
consistently predict correctly more than 60\% of the time.

Figure \ref{fig:synch-prediction-accuracy} also shows that prediction
accuracies diminish as the number of neurons for a synchronized event
increases from 2 to 8. This is expected; the higher the threshold for
synchronization, the more the number individual neurons that have to
predicted correctly. Despite this, prediction accuracies decrease by
only 10\% at worst.

\begin{figure}[t]
\centering
{
\begin{minipage}[t]{0.48\textwidth}
\centering
\vspace{-4mm}
\epsfig{file=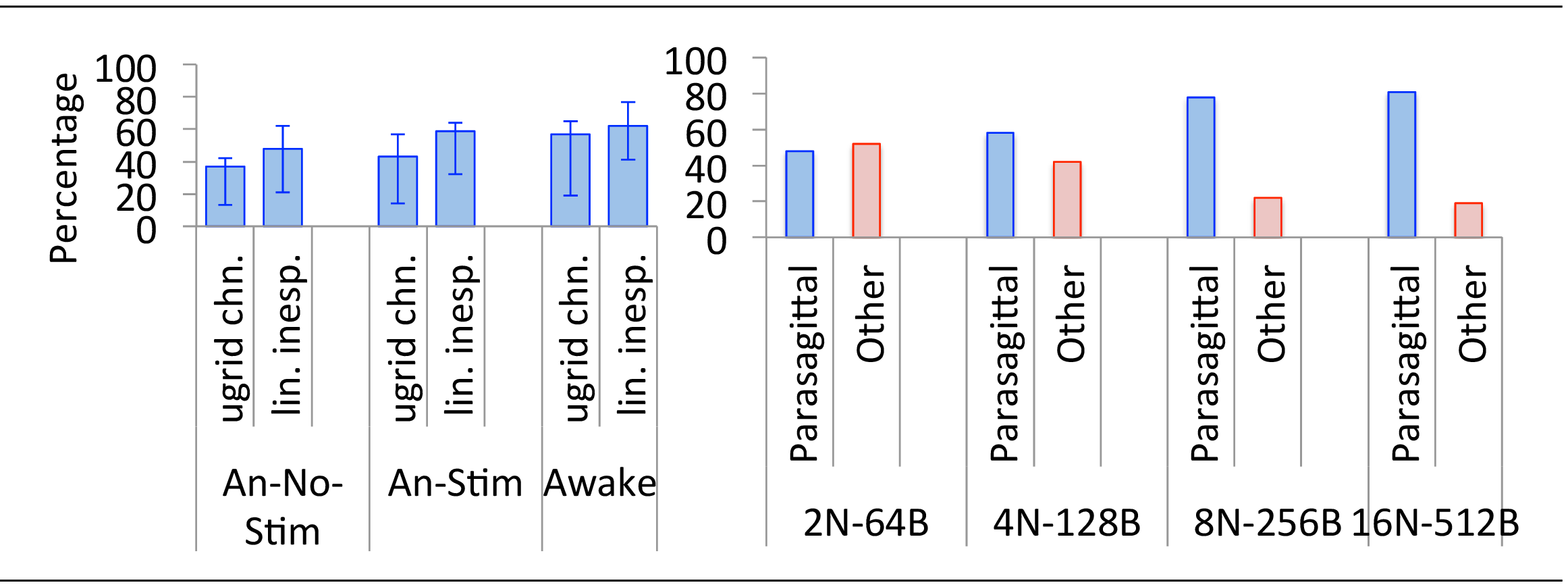, scale=0.36, angle=90}
\vspace{-48mm}
\caption{\small (Left) Percentage of Purkinje neurons that experience
  micro-grid changes ({\sf ugrid-chn.}) and are linearly inseperable
  ({\sf lin.-insep.}) with averages, min/max shown; and (right) for each
  neuron predicted to fire, percentage of total weighted sum value
  contributed by the weights from parasagittal neurons versus others.}
\vspace{-4mm}
\label{fig:linear-separability-parasagittal}
\end{minipage}
}
\end{figure}

\vspace{2mm}{\noindent \bf Understanding prediction characteristics:}
We now discuss the source of mispredictions, focusing on perceptrons
as they predict neuronal behavior most accurately. Like branch
misprediction, most neuronal misprediction aris\-es from neurons that
are linearly inseparable. Past work identifies the fraction of static
branches that are linearly inseparable to understand mispredictions
\cite{jimenez:perceptron}. The graph on the left in Figure
\ref{fig:linear-separability-parasagittal} does the same, but for
neuronal prediction ({\sf lin. insep.}). There is a biological basis
for linear inseparability -- neurons sometimes change which other
neurons they correlate with. We study our neuronal traces and every
10ms, identify micro-grids. As a fraction these samples, we plot the
percentage of time that neurons change between micro-grids ({\sf ugrid
  chn}). Figure \ref{fig:linear-separability-parasagittal} shows that
adding sensorimotor stimulation ({\sf An-Stim} and {\sf Awake})
increases micro-grid switches and linearly inseparable neurons,
lowering prediction accuracy.

The graph on the right in Figure
\ref{fig:linear-separability-parasagittal} shows that perceptron
predictors also accurately capture the biology of parasagittal
correlations. Every time the predictor predicts firing activity, we
log what percentage of the perceptron's weighted sum originates from
weights of parasagittal neurons. The higher the percentage, the higher
the correlations between parasagittal neurons. We plot these
percentages in blue (with the rest shown in red), as a function of the
perceptron predictor size and number of weights in each perceptron
entry (e.g., jN-kB indicates j weights and k bytes). As expected, with
more weights, parasagittal correlations are more easily tracked.

\vspace{2mm}{\noindent \bf Global versus global/local perceptron
  histories:} Beyond perceptrons with global history, we have also
studied a mix of local and global history
\cite{jimenez:neural}. Because prediction accuracy hinges on neuronal
correlations, we see little benefit (less than 1\% more accuracy) from
the addition of local history.

\begin{figure}[t]
\centering
{
\begin{minipage}[t]{0.48\textwidth}
\centering
\vspace{-4mm}
\epsfig{file=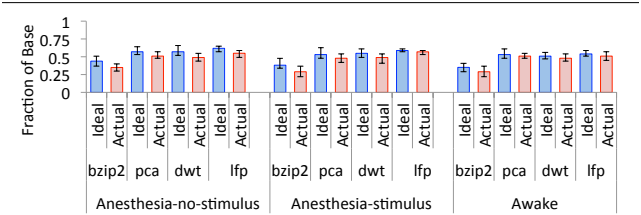, scale=0.36, angle=90}
\vspace{-48mm}
\caption{\small Fraction of baseline energy saved using {\sf Ideal}
  and {\sf Actual} prediction. We assume perceptrons with 32 8-bit
  weights (1KB budget) and that 4 neurons must fire to be considered
  synchronized.}
\vspace{-4mm}
\label{fig:detailed-energy}
\end{minipage}
}
\end{figure}

\begin{figure}[t]
\centering
{
\begin{minipage}[t]{0.48\textwidth}
\centering
\vspace{-4mm}
\epsfig{file=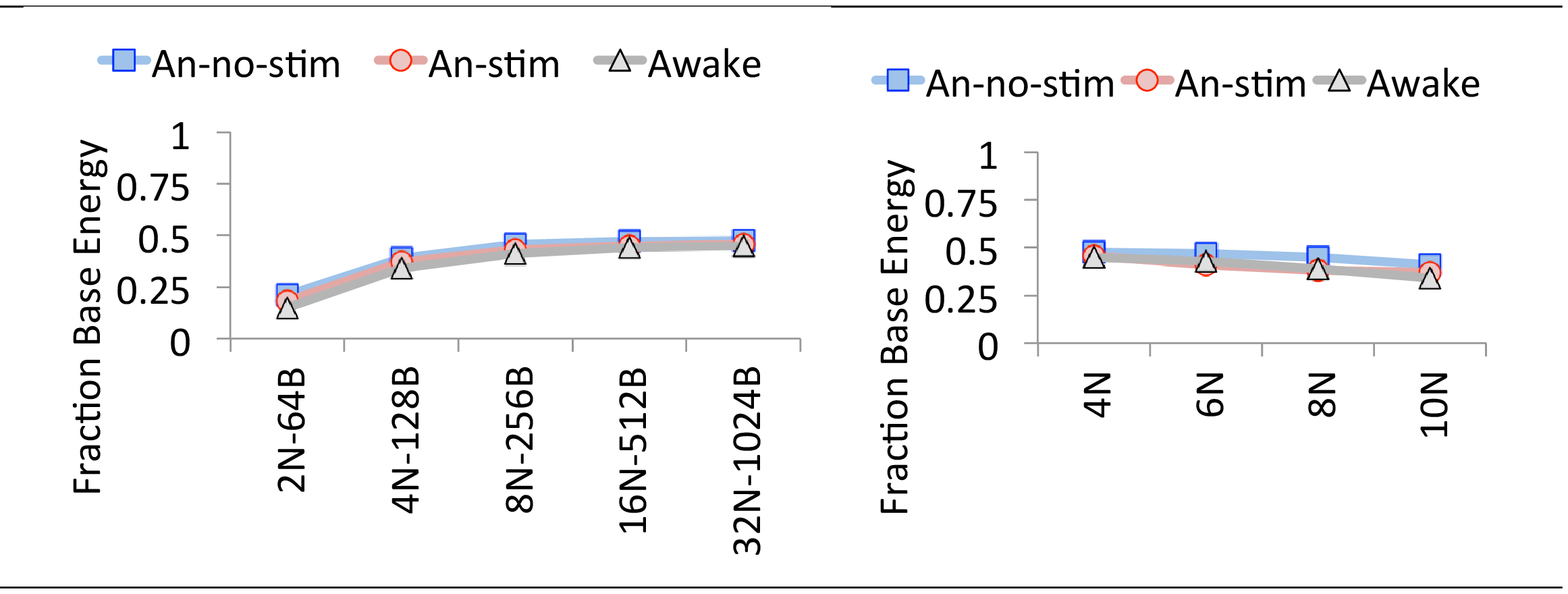, scale=0.36, angle=90}
\vspace{-48mm}
\caption{\small (Left) Average fraction of baseline energy saved when
  using a perceptron predictor, for different numbers of weights. We
  assume that 4 neurons must fire to be considered synchronized; and
  (right) average energy saved when using a perceptron predictor with
  32 8-bit weights (1KB total budget) and varying the number of
  neurons that must fire to be considered synchronized from 2 to 10.}
\vspace{-6mm}
\label{fig:energy-sensitivity}
\end{minipage}
}
\end{figure}

\vspace{2mm}{\noindent \bf Energy savings of perceptrons:} Figure
\ref{fig:detailed-energy} quantifies the fraction of energy saved
versus the baseline without neuronal prediction described in
Sec. \ref{sec:motivation}. The baseline uses an 8KB Smith branch
predictor, which we replace with other branch predictors. Though we
subsequently revisit the energy benefits of using Smith, gshare, and
two-level adaptive predictors for neuronal prediction, we focus on
perceptron predictors for now as their energy benefits far outweigh
the other approaches. Figure \ref{fig:detailed-energy} assumes 1KB
perceptron predictors, and that four neurons must fire close together
in time to be considered a synchronized event.

Figure \ref{fig:detailed-energy} shows that our energy savings ({\sf
  Actual}) achieve within 5-10\% of the {\sf Ideal} energy savings
from oracular neuronal prediction. Overall, this corresponds to energy
savings of 22-59\%. Naturally, applying stimuli to the mouse generally
decreases energy saving potential since there are more synchronized
events. Nevertheless, the neural implant still achieves 22-50\% energy
svaings on {\sf Awake} mice.

Figure \ref{fig:energy-sensitivity} sheds more light on energy
trends. The graph on the left shows the energy saved as the number of
8-bit weights per perceptron entry varies from 2 to 32. More weights
improve predictor accuracy by capturing micro-grid
correlations. Increasing the number of weights from 2 to 8 doubles
energy savings for anesthetized and awake mice. Meanwhile, the graph
on the right shows the average energy saved by a 1KB perceptron
predictor (with 32 weights per entry), as we vary the number of
neurons that must concurrently fire to be considered a synchronized
event. As the number of neurons increases, energy savings decrease as
there are fewer instances of synchronized events. Nevertheless, even
when we assume that 10 neurons must fire to constitute
synchronization, we save an average of 30\%+ of energy. And since
scientists generally study micro-grids of 4-8 neurons
\cite{ozden:reliable}, average energy savings are closer to 45\%+.

\vspace{2mm}{\noindent \bf Undesirable energy savings:} In
Sec.~\ref{sec:energy-management-strategies}, we illustrated situations
where the branch predictor predicts no synchronization, only to find
that it does occur. This loses important pre-synchronized activity, so
its energy savings are undesirable. We have quantified the percentage
of energy savings that are undesirable -- they are less than 2\% of
total energy savings for all workloads. The reason this number is
small is that perceptrons have good prediction accuracy. The (few)
mispredictions idle the processor for an extra 10ms (the time taken to
identify the misprediction). Subsequently, the processor transitions
to nominal operation. Compared to the long stretches of times that the
processor is correctly predicted and placed in idle low power mode
(10s of seconds), these mispredictions minimally affect energy saved.

\vspace{2mm}{\noindent \bf Energy savings versus other branch
  predictors:} We have focused on perceptron predictors since they
consistently save more energy than other approaches.  Smith, gshare,
and two-level adaptive predictors are simpler and enjoy lower access
energies, but their lower prediction rates means that they save, on
average, 5-45\% less energy than perceptrons.

\vspace{2mm}{\noindent \bf Impact of branch prediction on program
  behavior:} Of course, the choice of branch predictor also impacts
the runtime of the four workloads in Sec. \ref{sec:methodology}. While
sophisticated predictors can consume higher access energy, they may
also save overall energy if they cut down runtime sufficiently. We
have found that while perceptron branch predictors do increase energy
usage when the processor is in nominal operating mode (compared to
Smith and gshare prediction), they still save considerably more energy
overall because the potential energy savings from neuronal prediction
far outweigh any energy impact of performing branch prediction.

\vspace{2mm}{\noindent \bf Energy savings with dynamic
  voltage/frequency scaling:} We have focused on idle rather than
active low power modes like dynamic voltage/fre\-quency scaling
(DVFS). This is because Cortex M processors currently support only the
former. However, we were curious about energy savings if DVFS were to
be incorporated. Therefore, we studied and compared three schemes
using a 1KB perceptron predictor: (1) use idle low power modes as
described thus far; (2) use DVFS instead of neuronal prediction, to
identify opportunities when the workloads can afford to be slowed down
to 0.5$\times$ and 0.75$\times$ of the baseline frequency using
standard CPU utilization based prior approaches \cite{deng:coscale};
and (3) combine (1) and (2) by showing that DVFS and idle mode usage
with neuronal prediction are orthogonal. We found that (2) remains
superior to (1), saving an average of 12\% more energy. Combining
neuronal prediction and idle low power modes with DVFS promises even
more energy savings (as high as 67\%, exceeding the 59\% of neuronal
prediction alone). 

\vspace{-4mm}
\section{Concluding Remarks}\label{sec:discussion}
{\noindent \bf Generality of observations:} How
conclusively can we say that branch predictors, or indeed any hardware
predictors, can predict brain activity? To answer this question, we
need to study much more than just 26 minutes of neuronal activity,
from more locations than just lobule 6 of the cerebellum. Our study is
but a first step in this direction.


\vspace{2mm}{\noindent \bf Alternate predictors:} Perceptron
predictors cannot accurately predict brain activity in some cases. A
promising direction may be to study hardware that implements richer
machine learning techniques to predict dynamic program behavior, like
reuse distances of cache lines \cite{teran:perceptron}. Perhaps it may
even be possible to co-opt more sophisticated hardware neural networks
currently being studied \cite{sharma:from, chen:eyeriss}, or rely on
intelligent software machine learning techniques.

\vspace{2mm}{\noindent \bf Modeling brain circuits:} Neuroscientists
are activity seeking to model neural circuits that explain neuronal
biology \cite{joshi:integrated}. It may be fruitful to consider
whether models of well-known microarchitectural structures like branch
predictors could aid neural circuit modeling, particularly if the
microarchitectural structures predict neuronal activity accurately.

\vspace{2mm}{\noindent \bf Related work:} Our work is related to
hardware neural network accelerators \cite{sharma:from, chen:eyeriss,
  shen:maximizing, alwani:fused, liu:pudianno, chen:dadiannao,
  chen:high}, but is closer to studies that link neural biology with
computer architecture. For example, Hashmi et.~al.~studied fault
tolerance in cortical microarchitectures \cite{hashmi:automatic} while
Nere et.~al.~emulated biological neurons digitally
\cite{nere:bridging}. Their work paved the way for Smith's pioneering
studies on efficient digital neurons for large-scale cortical
architectures \cite{smith:efficient}. We are inspired by these studies
but focus on co-opting {\it existing} microarchitectural structures to
predict neuronal activity and manage power/energy.

\newpage
\bibliographystyle{ieeetr} \bibliography{references}

\end{document}